\documentclass[times,twocolumn]{aastex631}

\usepackage{natbib}
\usepackage{amsmath}
\usepackage{amssymb}
\usepackage{subfigure}
\usepackage{url}
\usepackage{CJK}
\usepackage{longtable}
\usepackage{graphicx}   
\usepackage{epsfig}
\usepackage{bm}
\usepackage{multirow}

\renewcommand{\vec}[1]{ {\mathbf #1} }

\newcommand{\Fig}{{Figure}}

\newcommand{\RNum}[1]{\uppercase\expandafter{\romannumeral #1\relax}}

\shorttitle{A fundamental mechanism of solar eruption initiation in multipolar magnetic field}
\shortauthors{Bian et al.}
\graphicspath{{figures/}}

\begin{document}

\title{A fundamental mechanism of solar eruption initiation in multipolar magnetic field}
	
\correspondingauthor{Chaowei Jiang} \email{chaowei@hit.edu.cn}

\author[0000-0001-9189-1846]{Xinkai Bian}
\affiliation{Institute of Space Science and Applied Technology, 
	Harbin Institute of Technology, Shenzhen 518055, China}

\author[0000-0002-7018-6862]{Chaowei Jiang}
\affiliation{Institute of Space Science and Applied Technology, 
	Harbin Institute of Technology, Shenzhen 518055, China}

\author[0000-0001-8605-2159]{Xueshang Feng}
\affiliation{Institute of Space Science and Applied Technology,
	 Harbin Institute of Technology, Shenzhen 518055, China}

\author[0000-0003-4711-0306]{Pingbing Zuo}
\affiliation{Institute of Space Science and Applied Technology,
	Harbin Institute of Technology, Shenzhen 518055, China}

\author[0000-0002-7094-9794]{Yi Wang}
\affiliation{Institute of Space Science and Applied Technology,
	Harbin Institute of Technology, Shenzhen 518055, China}

\begin{abstract}
Recently we established a fundamental mechanism of solar eruption initiation, in which an eruption can be initiated from a bipolar field through magnetic reconnection in the current sheet (CS) that is formed slowly in the core field as driven by photospheric shearing motion. Here using a series of fully 3D MHD simulations with a range of different photospheric magnetic flux distributions, we extended this fundamental mechanism to the quadrupolar magnetic field containing a null point above the core field, which is the basic configuration of the classical breakout model. As is commonly believed, in such multipolar configuration, the reconnection triggered in the CS originated at the null point (namely, the breakout reconnection) plays the key role in eruption initiation by establishing a positive feedback-loop between the breakout reconnection and the expansion of the core field. However, our simulation showed that the key of eruption initiation in such multipolar configuration remains to be the slow formation of the CS in the sheared core rather than the onset of fast breakout reconnection. The breakout reconnection only helps the formation of the core CS by letting the core field expand faster, but the eruption cannot occur when the bottom surface driving is stopped well before the core CS is formed, even though the fast reconnection has already been triggered in the breakout CS.  
This study clarified the role of breakout reconnection and confirmed formation of the core CS as the key to the eruption initiation in a multipolar magnetic field. 
\end{abstract}

\section{Introduction} \label{sec:intro}

Coronal mass ejections (CMEs) are the most spectacular eruptive phenomenon in the solar atmosphere, and its energy is derived from the free magnetic energy stored in the coronal magnetic field.
Due to photospheric line-tied effect, the coronal magnetic field is stressed and deviates from the potential field under the action of various photospheric motions (such as shear and rotational flows), during which free magnetic energy accumulates.
This evolution process is quasi-static, and the coronal magnetic field is in an approximately force-free state, that is, the outward magnetic pressure of the low-lying flux is balanced with the inward magnetic tension force of the overlying flux.   
At a critical point, this force equilibrium is disrupted and the eruption begins suddenly, during which the free magnetic energy is rapidly converted into impulsive heating and fast acceleration within the plasma.
However, how the balance of forces is disrupted, that is, the initiation mechanism of solar eruption, remains an open question.

Earlier studies generally suggested that CMEs arise from regions where the magnetic field is closed, while the occurrence of CMEs needs these closed magnetic structures to fully open. 
However, according to the Aly-Sturrock conjecture~\citep{Aly1991, Sturrock1991}, the energy of a fully open field is the upper limit of the energy of all possible force-free fields with a given magnetic flux distribution on the bottom and a simply connected topology, so CME cannot occur from the release of the magnetic energy, which is known as the Aly-Sturrock paradox. 
Therefore, all the theories of CME initiation need to avoid this paradox in some way~\citep{Forbes2006, Shibata2011, Chen2011, Schmieder2013, Aulanier2014, Janvier2015}, such as that the magnetic field before eruption is not simply connected, i.e., there is a pre-existing magnetic flux rope (MFR), or that the magnetic structure after eruption is not fully open (i.e., partially open) and magnetic reconnection plays a key role in the eruption.

For the models based on magnetic reconnection, many efforts have been made in early simulations of solar eruption initiated from the simplest magnetic configuration, i.e., a bipolar magnetic field, in two-dimensional (2D) or transformation-invariant coordinate systems~\citep{Choe1996, Mikic1994, Amari1996a}.
All these simulations show that by continuously applying shear to the bipolar field, the magnetic structure tends to approach the open field with a current sheet (CS) being formed over the polarity inversion line (PIL). Once the magnetic resistivity is applied, magnetic reconnection immediately sets in at the CS and initiates the eruption.
Unfortunately, such a simple scenario failed to work in any fully three-dimensional (3D) simulation at that time. Moreover, in the 2D cases, the eruption requires full open of the overlying field, which is inconsistent with the observation. 
As such, \citet{Antiochos1999} proposed an alternative scenario, called as the breakout mechanism, based on a multipolar field, in which only the core flux erupts and neighboring flux is still closed. This mechanism requires a quadrupolar magnetic field consisting of a compact core bipolar field and a background bipolar field of opposite orientation to that of the core field.
Then magnetic null point naturally forms above the core field. Through continuous shearing of the core field, which energizing the system, the core field expands and squeezes the magnetic null point to form a CS (referred to as the breakout CS). When the finite resistivity is considered, magnetic reconnection will occur and remove the un-sheared, overlying flux above the low-lying, sheared core flux, thus allowing the core field to further expand and enhance the breakout CS. This positive feedback loop eventually causes the eruption.

There have been many developments since the breakout mechanism was proposed~\citep{MacNeice2004, Amari2007, Lynch2009, Pariat2010, Wyper2016, Dahlin2019, Kumar2021}. 
For example, \citet{Lynch2008} reproduced the process of the breakout mechanism initiating eruption using 3D MHD simulation, and \citet{DeVore2008} demonstrated that homologous eruptions could be triggered by the breakout mechanism, although without producing CMEs. 
Also, a well-observed eruption was found to be in good qualitative agreement with the topology and dynamical evolution of the breakout mechanism~\citep{Chen2016}.
This mechanism is also applicable to jet-related eruptions, suggesting that breakout mechanism may be a universal model for solar eruption from large scale CMEs to small-size eruptive activities~\citep{Wyper2017}. 
In the ultra-high resolution 2D MHD simulation of the breakout mechanism~\citep{Karpen2012}, it was further found that the start of fast breakout reconnection (indicated by Alfv$\acute{\text{e}}$nic outflows in the breakout reconnection), rather than the initial reconnection at the null point, corresponds to the eruption onset. Meanwhile, it is found that the explosive release of magnetic energy and the rapid acceleration of CME are caused by the reconnection in the flare CS which is the central CS formed in the core field during its dynamic expansion as driven by the breakout reconnection.

In our recent works based on high-accuracy, fully 3D MHD simulations~\citep{Jiang2021b, Bian2022a, Bian2022, Bian2022b}, we for the first time found that in the absence of background bipolar field, that is, without the magnetic null point and the breakout reconnection, solar eruption can still occur in the simply connected bipolar field due to the reconnection of the core CS that is formed slowly as driven by photospheric shearing motion, and we suggest this model to be a fundamental mechanism of solar eruption due to its simplicity and efficacy. 
Then the question naturally arises, since applying shearing flow in a single bipolar field can form a core CS and produce an eruption, what effect does the addition of a background field (that forms a magnetic null point) have on the initiation of eruption? and whether the fast breakout reconnection can be used a sign of the eruption onset?

In order to answer these two questions, here we conducted a series of fully 3D MHD simulations with the same core bipolar field but different background bipolar field, so they have different location of magnetic null point. Meanwhile, we also conducted simulation without the background field for comparison. 
We found that all simulations with continuously bottom driving (for energizing the core field) produced an eruption, and the onset of the eruption is related to the initial height of null point, namely, the smaller the height is, the earlier the eruption starts. 
This indicates that the breakout reconnection indeed helps the formation of the core CS by letting the core field expand faster. 
But the breakout reconnection releases very little of the magnetic energy, and the rapid release of magnetic energy only corresponds to the reconnection of the core CS. In addition, we selected some simulations and stopped driving at certain moments when the fast breakout reconnection has started while the core CS has not been formed yet. 
These experiments show that there exists a critical point in time which may correspond to the transition from a quasi-static evolution phase to a slow-rise phase prior to the eruption onset. Before the critical point, the system cannot produce eruption without the bottom driving, because the core CS fails to form, even though the fast breakout reconnection has been triggered.
In such case, the super-Alfv$\acute{\text{e}}$nic velocity provided by the fast breakout reconnection will be dissipated with the relaxation of the system to a stable state, and cannot lead to a feedback loop between the breakout reconnection and the expansion of the core field. 
On the other hand, when the system has passed the critical point, the eruption is inevitable since the core CS can spontaneously form without the bottom driving. Such behavior exists in both the single bipolar field and the quadrupolar field, and thus is not dependent on the breakout reconnection.
This shows that the formation of core CS is the key to eruption. Therefore, the fundamental mechanism can be extended to the multipolar magnetic fields.

This paper is organized as follows. In Sect.~\ref{sec:model}, we define the magnetograms for the quadrupolar field with different background bipolar field, and briefly introduce our MHD model.
Then we show the results of five simulations with continuously bottom driving and analyze the role of breakout reconnection in Sect.~\ref{subsec:erupt} and \ref{subsec:comparison}.
The simulation and analysis revealing the key of the eruption are given in Sect.~\ref{subsec:key}.
Then, the analysis of the slow rise phase is presented in Sect.~\ref{subsec:slow_rise}.
Finally, our conclusion and discussion are given in Sect.~\ref{sec:con}.

\section{MHD model} \label{sec:model}
We numerically solve the full MHD equations in 3D Cartesian coordinate to study the dynamic evolution of solar corona. The MHD solver  is based on the conservation element and solution element (CESE) method and uses adaptive mesh refinement (AMR) grid~\citep{Feng2010, Jiang2010, Jiang2016a, Jiang2021b}. 
Since the controlling equations and the numerical code are the same as used in \citet{Jiang2021b}, we will not repeat them here, except some key settings and parameters as follows. 

Our model includes solar gravity and plasma pressure in the momentum equation, and the initial plasma density distribution satisfies hydro static equilibrium.
Meanwhile, a small viscosity $\nu=0.05\frac{(\Delta x)^2}{\Delta t}$ (where $\Delta x$ and $\Delta t$ are the local grid resolution and time step, respectively) is used in the momentum equation to keep numerical stability during the dynamic phase of the simulated eruptions.
There is no explicit resistivity used in the magnetic induction equation, but magnetic reconnection can still occur due to numerical resistivity when the thickness of a current layer is close to the local grid resolution.

\subsection{Boundary Condition}
In order to comprehensively understand the role of the magnetic null point (breakout reconnection) in producing eruption, we conducted a series of MHD simulations using photospheric magnetograms with the same core bipolar field but with different background field, thus forming different quadrupolar configurations. 

\begin{figure*}[!htbp]
	\centering
	\includegraphics[width=0.8\textwidth]{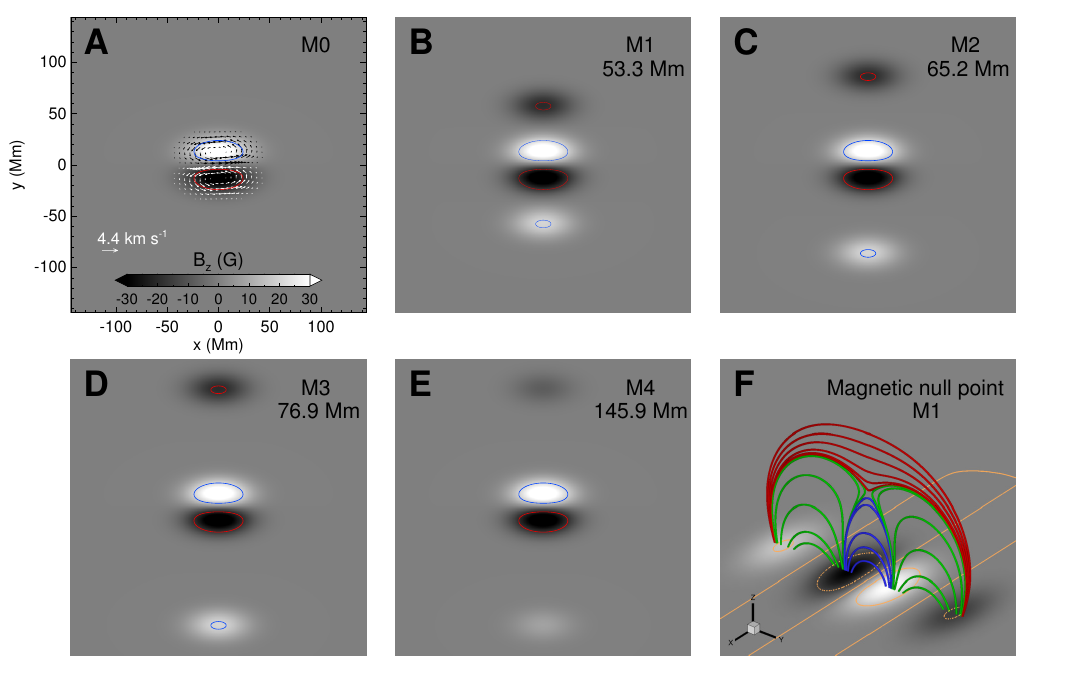}
	\caption{Magnetograms used in all the different simulations and a schematic representation of the magnetic null point. \textbf{A}, Magnetic flux distribution and surface rotation flow at the bottom boundary (i.e., $z=0$). The background is color-coded by the vertical magnetic component $B_{z}$, and the vectors show the rotational flow. The contour lines of the blue and red lines represent $\frac{1}{2}$ of the maximum and minimum values of the vertical magnetic  components $B_{z}$ on the bottom, respectively. The formats of panels~\textbf{B}, \textbf{C}, \textbf{D} and \textbf{E} are the same as those of \textbf{A}, but without showing the rotational flow. The name of the magnetograms and the height of the magnetic null point are also marked in each panel.
	\textbf{E}, Potential field structure obtained by extrapolation using magnetogram M1. The thick curves with different colors represent magnetic field lines belonging to different flux systems, which meet at the $X$ point, i.e., the magnetic null point. The yellow thin lines are contours of $B_z$.}
	\label{fig:map}
\end{figure*}

At the bottom surface (i.e., $z=0$ plane), the core bipolar field is defined as the composition of two Gaussian functions~\citep{Amari2003, Jiang2021b},
\begin{equation}\label{eq:core_field}
	\begin{split}
		B_{z,{\rm core}}(x,y,0) = B_0 e^{-x^2/\sigma_x^2}(e^{-(y-y_c)^2/\sigma_y^2}\\-e^{-(y+y_c)^2/\sigma_y^2}),
	\end{split}
\end{equation} 
and the background bipolar field is defined as,
\begin{equation}\label{eq:back_field}
	\begin{split}
		B_{z,{\rm back}}(x,y,0) = -\epsilon B_0 e^{-x^2/\sigma_x^2}(e^{-(y-y_b)^2/\sigma_y^2}\\-e^{-(y+y_b)^2/\sigma_y^2}),
	\end{split}
\end{equation}
where $B_0$ is a constant such that the maximum value of photospheric $B_z$ is $37.2$~G. The parameters $\sigma_x$ and $\sigma_y$ are $28.8$~Mm and $14.4$~Mm, respectively, which control the magnetic flux distribution range in the $x$ and $y$ direction.
The parameter $y_c$ controls the distance between the positive and negative poles of the core field, with a value of $11.52$~Mm, and the parameter $y_b$ controls the distance between the positive and negative poles of the background field.
The parameter $\epsilon$ is the ratio of the maximum $B_z$ of the background field to that of the core field on the bottom.
By setting different $y_b$ and $\epsilon$, we can obtain a series of different distributions of background bipolar field while their core field are the same. 
As a result, the quadrupolar configurations have different locations of null point and different ratios of the background magnetic flux to the core flux. 
Table~\ref{tab:Map_Expression} gives the specific parameter settings. These magnetograms used in the simulation are shown in \Fig~\ref{fig:map}, and will be referred to as M0 to M4.

\begin{table}[!htbp]
	\centering
	\begin{tabular}{ccccc}	
		\hline\hline
		\multirow{2}{*}{Map}&\multicolumn{2}{c}{Parameters}& \multirow{2}{*}{$\frac{\Phi_{\rm back}}{\Phi_{\rm core}}$}& Height of magnetic \\ 	
		\cline{2-3}
		&$y_b$ (Mm) & $\epsilon$ & & null point (Mm) \\ 
		\hline
		M0 & 0     & 0   & 0    & ---  \\
		M1 & 57.6  & 0.5 & 0.67 & 53.3      \\
		M2 & 86.4  & 0.5 & 0.67 & 65.2      \\
		M3 & 115.2 & 0.5 & 0.67 & 76.9        \\
		M4 & 115.2 & 0.2 & 0.27 & 145.9       \\
		\hline
	\end{tabular}
	\caption{Parameters $y_b$ and $\epsilon$ that defines the five magnetograms of M0 to M4. In the table, the ratio of the total unsigned flux of the background field to the core field and the height of the magnetic null point are also given.}
	\label{tab:Map_Expression}
\end{table}

Our series of MHD simulations all begin with the potential field which is computed using the Green’s function method. Therefore, the initial potential field has a magnetic null point above the core field PIL, and its height can be obtained, as shown in \Fig~\ref{fig:map}F and Table~\ref{tab:Map_Expression}. 
Since the parameter $\epsilon$ is set to be less than $1$, the total unsigned flux of the background field is smaller than that of the core field, which is often the case in the realistic solar active regions. Meanwhile, the distance of the background bipolar field also affects the height of the magnetic null point, namely, the smaller the distance is, the lower the null point resides.

To add free magnetic energy to the system, the simulations are driven by rotational flow applied at the footpoints of the core field, which creates magnetic shear along the core PIL and does not modify the flux distribution on the bottom, as shown in \Fig~\ref{fig:map}A. The rotational flow is defined as 
\begin{equation}\label{eq:dirven_speed}
	\begin{split}
		v_{x}=\dfrac{\partial \psi(B_{z,{\rm core}})}{\partial y}; v_{y}=\dfrac{\partial \psi(B_{z,{\rm core}})}{\partial x}
	\end{split}
	,\end{equation}
with $\psi$ given by
\begin{equation}\label{eq:dirven_speed_psi}
	\begin{split}
		\psi = v_{0}B_{z,{\rm core}}^{2}e^{-(B_{z,{\rm core}}^{2}-B_{z, {\rm max}}^{2})/B_{z, {\rm max}}^{2}},
	\end{split}
\end{equation}
where $B_{z,{\rm max}}$ is the maximum value of the photospheric vertical magnetic component $B_{z}$, and $v_{0}$ is a constant for scaling such that the maximum of the surface velocity is $4.4$~km~s$^{-1}$, close to the typical flow speed in the photosphere ($\sim$$1$~km~s$^{-1}$).
The flow speed is smaller than the sound speed ($110$~km~s$^{-1}$) by two orders of magnitude and the local Alfv$\acute{\text{e}}$n speed (the largest Alfv$\acute{\text{e}}$n speed is $2300$~km~s$^{-1}$) by three orders, respectively, thus representing a quasi-static stress of the coronal magnetic field.

\subsection{Grid Setting}\label{grid}
The computational domain is large enough for simulating the eruption initiation process, spanning a Cartesian box of $(-270, -270, 0)$~Mm $\leq$ $(x, y, z)$ $\leq$ $(270, 270, 540)$~Mm. The full volume is resolved by a block-structured grid with AMR in which the base resolution is $2.88$~Mm, and the highest resolution of $360$~km is used to capture the formation of CSs and the subsequent reconnections. 
Besides capturing the CS and reconnection in the core field as done in \citet{Jiang2021b}, the grid refinement is controlled by the following criteria to ensure that the formation of the breakout CS and the subsequent reconnection were resolved using the highest resolution,
\begin{equation}\label{eq:refine}
  \begin{cases}
    B < 0.1, \\
	\dfrac{J}{B} > \dfrac{0.1}{\delta}, \\
	\nabla\rho<0.0001,
  \end{cases}
\end{equation}
where $\delta$ is the length of the local grid. The first two conditions are used to locate the signature of breakout CS, and the third condition is used to exclude unnecessary grid refinement of shock due to super-Alfv$\acute{\text{e}}$nic velocity.
If all the three conditions in Eq.~\ref{eq:refine} are met, the grid will be refined. 
Note that all quantities used here are expressed in their normalized values with units of magnetic field as $1.86$~G, length $11.52$~Mm, and density $2.29\times10^{-15}$~g cm$^{-3}$, respectively.

\section{Results} \label{sec:res}
We conducted a series of simulations using five different magnetograms, and carried out a comparative analysis of these different simulations. We first briefly described the evolution of a typical eruption and then analyzed the role of breakout reconnection by comparing the results of five continuously-driven simulations.
Furthermore, we selected three simulations to stop the bottom driving at three different moments and let the system evolve spontaneously to find the key factor determining whether the eruption can occur.
Finally, we use an artificial frictional force in some driving-stopped simulations to constrain the velocity of the system (thus can reduce the effect of inertia) and conduct analysis of the slow-rise phase. 
In total, twenty-two sets of simulations have been carried out in this study.

\subsection{A typical evolution} \label{subsec:erupt}
\begin{figure*}[!htbp]
	\centering
	\includegraphics[width=0.8\textwidth]{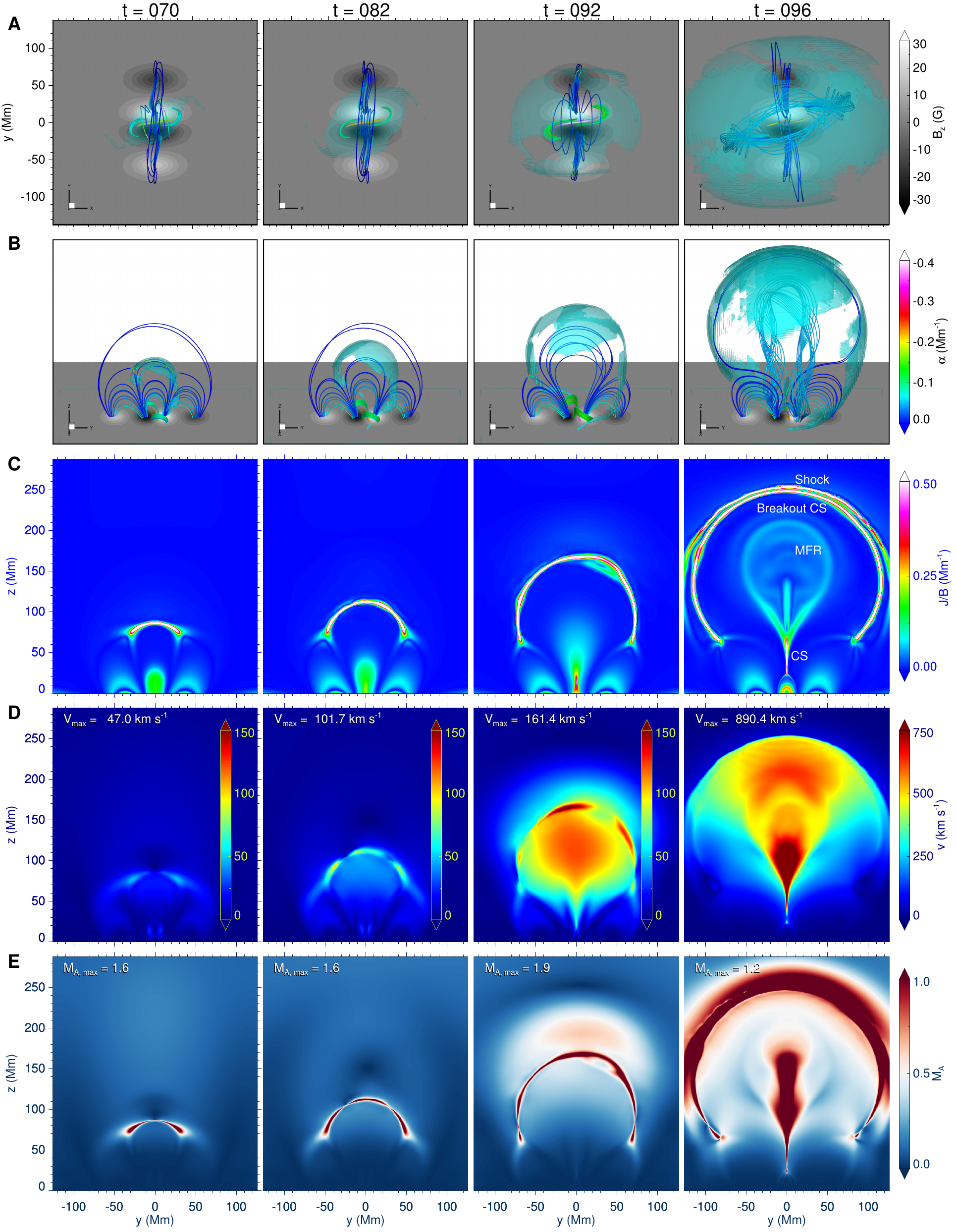}
	\caption{Evolution of magnetic field lines, structure of breakout CS, electric current, velocity and Alfv$\acute{\text{e}}$nic Mach number during the simulation M1. 
	\textbf{A}, Top view of magnetic field lines and the structure of breakout CS. The colored lines represents the magnetic field lines, and the colors denote the value of the nonlinear force-free factor. The cyan iso-surface represent the current sheet with $J/B=0.5$.		
    \textbf{B}, 3D perspective view of magnetic field lines and the structure of breakout CS. \textbf{C}, The central vertical slice (i.e., the $x=0$ plane) of current density $J$ normalized by magnetic field strength $B$ at different times. \textbf{D}, Velocity in the central vertical slice. \textbf{E}, Alfv$\acute{\text{e}}$nic Mach number distribution in the central vertical slice. The maximum velocity and Alfv$\acute{\text{e}}$nic Mach number are also given. (An animation of this figure is available. The animation is only of panels A-D.)}
	\label{fig:Breakout_slice}
\end{figure*}

\Fig~\ref{fig:Breakout_slice} (and its animation) show the evolution of 3D magnetic field lines, breakout CS, and the vertical cross section of the current density, velocity and Alfv$\acute{\text{e}}$nic Mach number in the continuously-driven simulation with magnetogram M1. 
After a period of surface rotational flow, the magnetic structure has a significant expansion, evolving from the initial potential field with magnetic null point to a configuration with strong shear above the core PIL and with the breakout CS formed. 
The 3D structure of breakout CS appears like a ``paraglider'' and in the 2D cross section it shows an arc on top of the core field. 
At $t = 70$ (the time unit is $\tau = 105$ s), with ongoing of the reconnection in the breakout CS, super Alfv$\acute{\text{e}}$nic velocity flow appeared around the breakout CS, and the maximum speed has reached $47$~km~s$^{-1}$ and the Alfv$\acute{\text{e}}$n Mach number of $1.6$.
However, in the core field below the breakout CS, it is still a quasi-static evolution since the speed is far smaller than the local Aflv$\acute{\text{e}}$n speed. Also in the core field region, there is no CS structure, but a volumetric current distribution.
With the further injection of energy from the bottom surface, the system expands more and more, and the breakout CS is squeezed to grow continually and plasmoid instability is triggered there, which leads to the reconnection in a turbulent way.
Meanwhile, the core volumetric current was squeezed to a current layer and gradually formed a vertical CS until $t=93$. The continuous increase of the energy in the system shown in \Fig~\ref{fig:energy_all}A (red line) is consistent with the energy injected continuously from the bottom surface. During this process, the total kinetic energy remains at a very low level. Even the onset of the fast breakout reconnection does not change the magnetic and kinetic energies significantly.

\begin{figure*}[!htbp]
	\centering
	\includegraphics[width=0.8\textwidth]{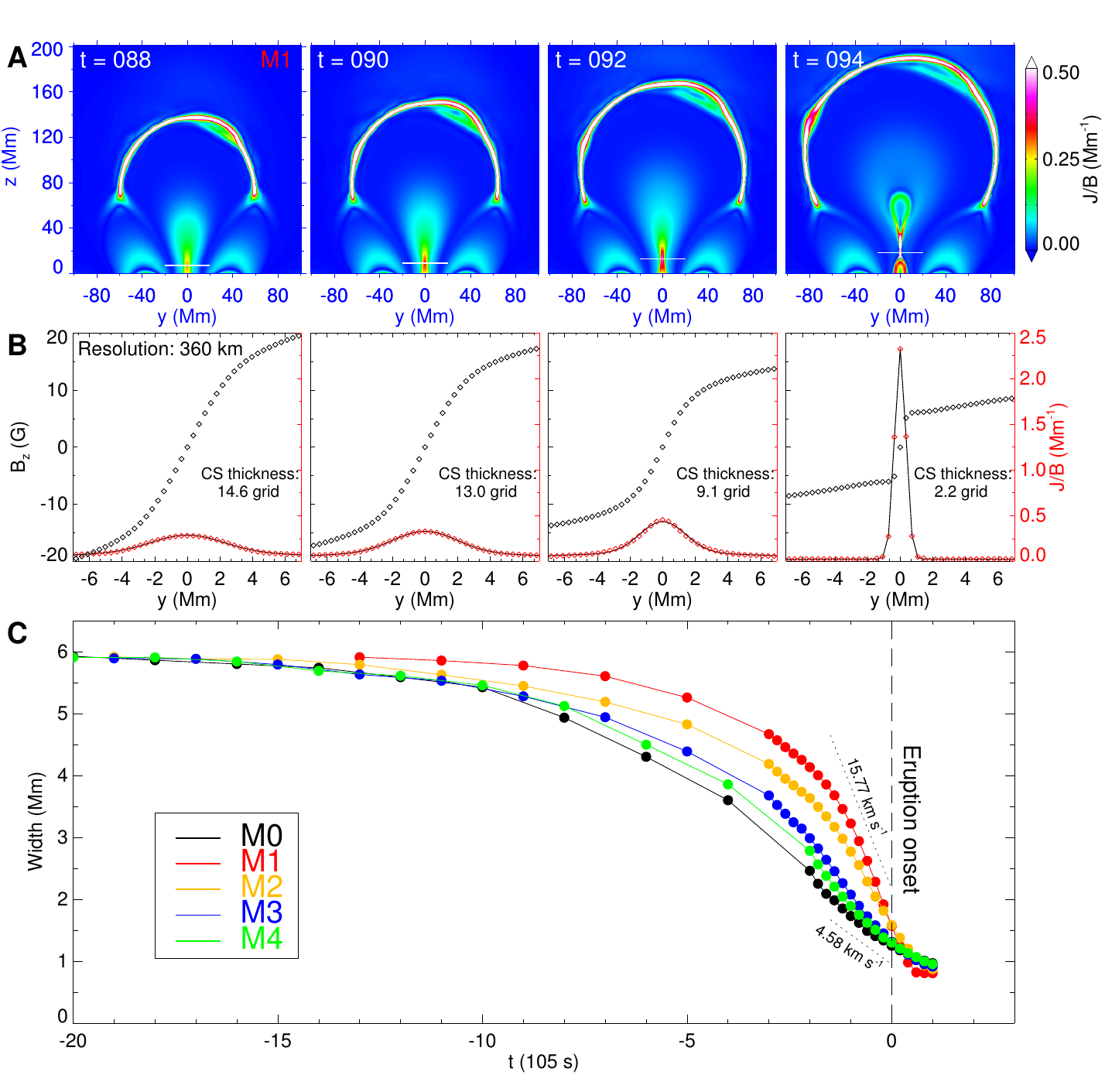}
	\caption{Formation process of the core CS. \textbf{A}, Distribution of normalized current $J/B$ on the $x=0$ slice in the simulation M1. \textbf{B}, One dimensional profile of the magnetic component $B_z$ and normalized current density along the horizontal line perpendicular to the center of the core CS (location with the maximum $J/B$). The diamonds represents the value on the grid nodes. The thin black line is a Gaussian function fitting of the normalized current density $J/B$ profile, and its FWHM is used to measure the thickness of CS.
    \textbf{C}, Temporal evolution of the thickness of the core CS in five simulations. The time axis of all the simulations are shifted so that $t=0$ denoted the onset of eruption, as shown by the vertical dashed line. Each circle represents a moment of output data, and a higher time resolution is used near the eruption onset. 
    The maximum speed of the core CS thinning in M1 and M0 are also denoted by the dotted lines and the numbers.}
	\label{fig:Jc_core_all}
\end{figure*}

The key transition occurs at $t=93$, when the thickness of the core current layer decreases to close to the grid resolution (\Fig~\ref{fig:Jc_core_all}), then magnetic reconnection kicks in there, resulting in rapid release of magnetic energy and sharp increase of kinetic energy. Obviously, that moment defines the eruption onset.

\begin{figure*}[!htbp]
	\centering
	\includegraphics[width=0.8\textwidth]{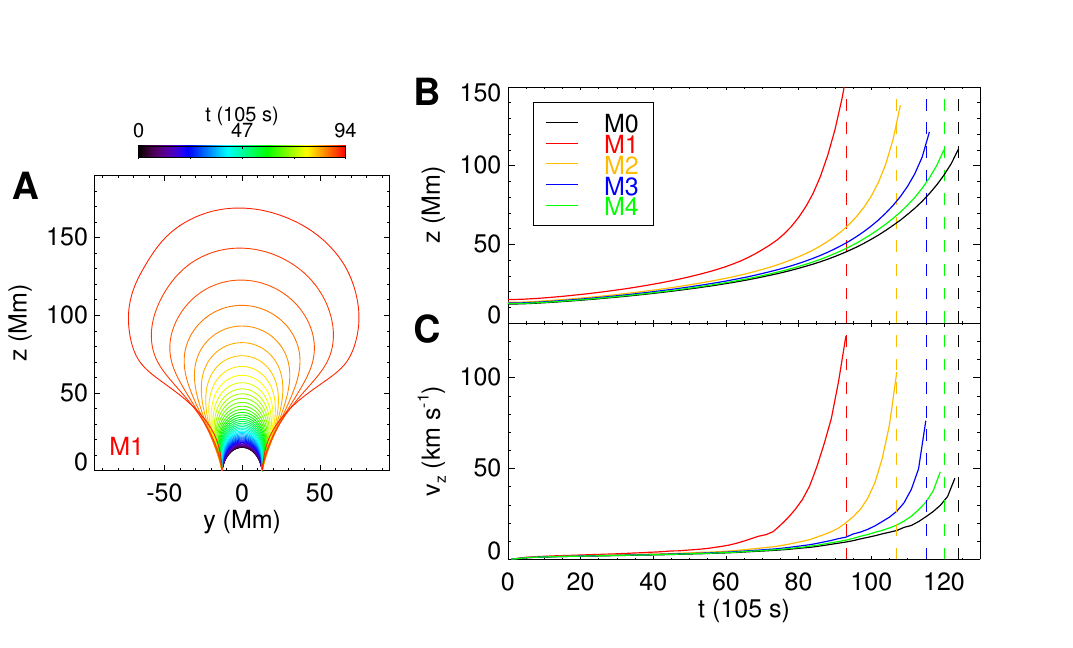}
	\caption{Pre-eruption evolution of the magnetic field line anchored at the center of the positive polarity of the core field. The rising of the apex of this field line is used to indicate the expansion of the core field. \textbf{A}, In simulation M1, the evolution of the projection of the magnetic field line on the $x=0$ plane. 	
	\textbf{B}, The apex of the magnetic field line varies with time in five simulations. The vertical dashed lines are shown for denoting the onsets of the eruption. \textbf{C}, The rising speed of the apex in the five simulations.}
	\label{fig:fieldlines}
\end{figure*}

After the eruption starts, the sheared magnetic arcades of the core field form a complex MFR through the reconnection of the core CS, which rapidly expands and rises.
Due to the sharp rise of the MFR, a fast magnetosonic shock is formed in front of it, sweeps through the breakout CS, and finally reaches the outer edge of the entire explosive structure, as shown in \Fig~\ref{fig:Breakout_slice} and its animation.
The initial reconnection height of core CS is around $20$~Mm above the core PIL (\Fig~\ref{fig:Breakout_slice}A).
By tracing the temporal evolution of the field line that roots in the center of the core polarities, as shown in \Fig~\ref{fig:fieldlines}A, we find that the two groups of magnetic arcades with initial reconnection are lower than this magnetic field line, that is, the initial core reconnection occurs between the shear magnetic arcades of the core field.
As the eruption continues, this magnetic field line first undergoes the breakout reconnection above and becomes part of the neighbor field, and then undergoes flare reconnection into a flare loop, as shown in animation of \Fig~\ref{fig:Breakout_slice}.

The simulation with magnetogram M1 show that overall the global evolution of the system agrees with the breakout model in respect of magnetic topology.

\subsection{The role of breakout reconnection in eruption}
\label{subsec:comparison}
\begin{figure*}[!htbp]
	\centering
	\includegraphics[width=0.8\textwidth]{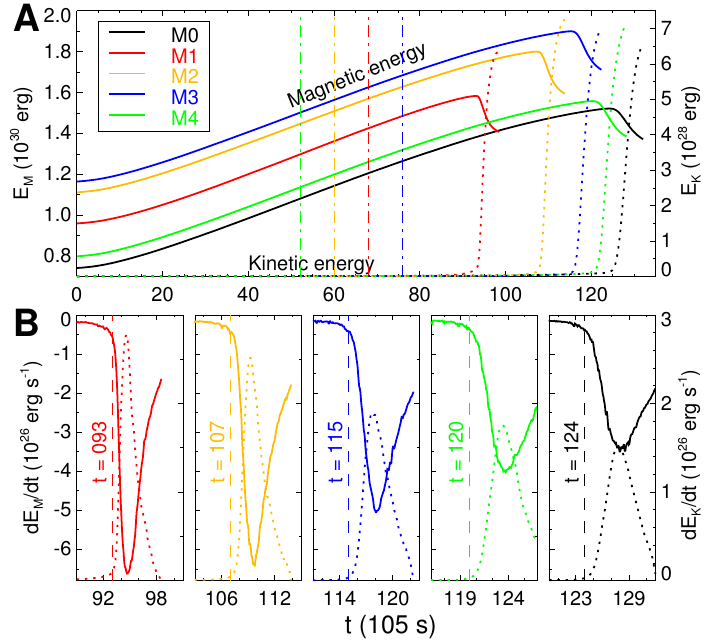}
	\caption{Temporal evolution of magnetic and kinetic energies in the five continuously-driven simulations. In all the panels, the colors of the curves (red, yellow, blue, green, and black) represent different simulations of M1, M2, M3, M4, and M0, respectively.
	\textbf{A}, Evolution of magnetic energy $E_{\rm M}$ (solid lines) and kinetic energy $E_{\rm K}$ (dotted lines). The vertical dashed dotted lines indicates the start moments of fast breakout reconnection in the different simulations.
	\textbf{B}, Releasing rate of magnetic energy and increasing rate of kinetic energy, with each subplot for each simulation. The vertical dashed lines show the onsets of the eruption, which are denoted on subplot. (An animation of this figure is available. The animation contains the same diagrams as panel C-D in \Fig~\ref{fig:Breakout_slice} and the evolution of kinetic energies in this \Fig.)}
	\label{fig:energy_all}
\end{figure*}

All five simulations with the different magnetogram from M0 to M4 produce eruption under continuously bottom driving, as shown in \Fig~\ref{fig:energy_all} (and its animation). 
The energy evolutions of these five simulations are similar, as shown in \Fig~\ref{fig:energy_all}A. The magnetic energies first increase almost linearly due to the continual injection of Poynting flux from the bottom surface, and then drop sharply at a certain time, that is, at the eruption onset.
Note that the initial magnetic energy (i.e., the potential field energy) differs in the different simulations because it is directly related to the different distributions of magnetic flux. However, since all magnetograms have the same core field and driving speed, their energy increase rates are equal, that is, these curves of magnetic energy are nearly parallel to one another before the eruption onset.
Even though the breakout reconnection occurred at different times in the four simulations (M1 to M4), the magnetic energy reduction was so small that it changes very limited in the energy evolution curve (compare with M0). Interestingly, the eruption onset time is clearly correlated with the initial height of the null point, as the lower the null point is, the earlier the eruption begins. The simulation with magnetogram M0 of the bipolar field can be regarded as with an infinitely far and small background field such that the null point is infinitely high, therefore it also satisfies the above relationship.

\Fig~\ref{fig:Jc_core_all} shows the temporal evolution of the thickness of core current layer in these five simulations. Note that for a better comparison, the times of the different simulations are shifted such that all the eruption onset times are $t=0$. 
All the simulations show that the thickness of core CS at the eruption onset is close to the grid resolution, of $3\sim 4$ grids at which the reconnection is triggered. This result clearly indicates that the reconnection of core CS results in the violent eruption in all the simulations. 
The thinning speed of the current layer is no more than $16$~km~s$^{-1}$, and is far below the local Alfv$\acute{\text{e}}$n speed (on the order of $1500$~km~s$^{-1}$).
Thus the formation of the CS is a quasi-static evolution, even though the outflow of fast breakout reconnection has reached $160$~km~s$^{-1}$ in simulation M1.
It can also be seen that thinning speed is related to the initial height of the null point, that is, the lower the null point, the greater the speed (and thus the less time to form CS).

\Fig~\ref{fig:fieldlines} shows the temporal evolution of the apex of the magnetic field line that is anchored at the positive polarity center of the core field in the five simulations. The rising of this field line indicates the expansion of the core field. As can be seen, the lower the null point is, the faster this magnetic field line rises, indicating that the whole core field expands faster.

Through the analysis and comparison of these simulations, we found that the magnetic null point or breakout reconnection indeed helps the eruption to occur. The reason is that an earlier and stronger breakout reconnection (due to a lower height of null point) would make the core field expand faster, thus the core CS can form earlier, and the reconnection of core CS immediately initiates the eruption.

\subsection{The key factor for initiation of eruption} 
\label{subsec:key}
To investigate the key factor determining whether an eruption can be initiated, we selected three simulations (M0, M1, and M3) to stop the bottom driving (i.e., the surface shearing motion) at three different moments before the eruption onset to see whether an eruption can still occur and analyze the reason. 
The three moments as selected to stop driving are, respectively, the first one (denoted by S1) is very close to the eruption onset (the core CS is almost formed), the second one (denoted by S2) is far away from the eruption onset (the core CS is just beginning to form), and the third one (denoted by S3) is even further ahead of the eruption onset. 
Note that for the simulations of quadrupolar field (i.e., M1 and M3), all the three moments of stopping driving are after the start of fast breakout reconnection. See Table~\ref{tab:simulation_S} for specific selection of simulations and moments.

\begin{table}[!htbp]
	\centering
	\begin{tabular}{ccccc}
		\hline\hline
		\multirow{2}{*}{Map} & Eruption  & \multirow{2}{*}{Start of FBR} & \multirow{2}{*}{Simulation} & Time \\
		& Onset & && (stop driving)  \\
		\hline
		\multirow{3}{*}{M0} & \multirow{3}{*}{123} & \multirow{3}{*}{---} & M0S1 & 120 \\
		&&&M0S2&116 \\
		&&&M0S3&112 \\
		\hline
		\multirow{3}{*}{M1} & \multirow{3}{*}{93} & \multirow{3}{*}{68} & M1S1 & 88 \\
		&&&M1S2&86 \\
		&&&M1S3&70 \\
		\hline
		\multirow{3}{*}{M3} & \multirow{3}{*}{114} & \multirow{3}{*}{76} & M3S1 & 110 \\
		&&&M3S2&104 \\
		&&&M3S3&100 \\
		\hline
	\end{tabular}
    \caption{The definition of nine driving-stopped simulations. The Start of FBR is the start time of fast breakout reconnection.}
	\label{tab:simulation_S}
\end{table}

\begin{figure*}[!htbp]
	\centering
	\includegraphics[width=0.8\textwidth]{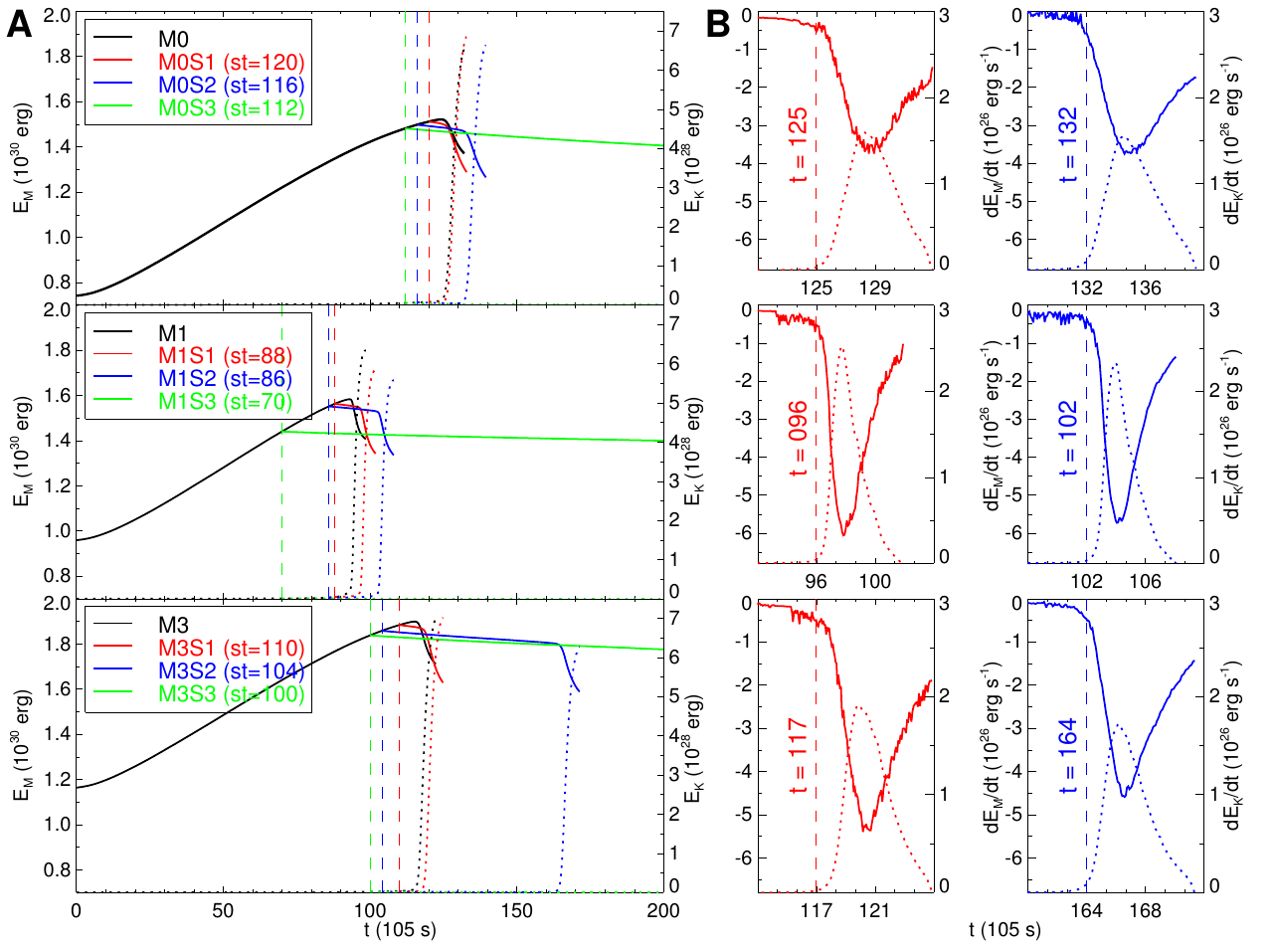}
	\caption{Temporal evolution of energies in nine driving-stopped simulations. \textbf{A}, Evolution of magnetic energy $E_{\rm M}$ (solid lines) and kinetic energy $E_{\rm K}$ (dotted lines). The red, blue and green in each panel denote the simulation with driving-stopped at different moments, called S1, S2, and S3, respectively. The vertical dashed lines are shown for denoting the moments when driving stops. \textbf{B}, Releasing rate of magnetic energy and increasing rate of kinetic energy of six simulations that produce eruption (S1 and S2). The vertical dashed lines with the time numbers show the onsets of the eruption.}
	\label{fig:energy_all_S}
\end{figure*}

\begin{figure*}[!htbp]
	\centering
	\includegraphics[width=0.8\textwidth]{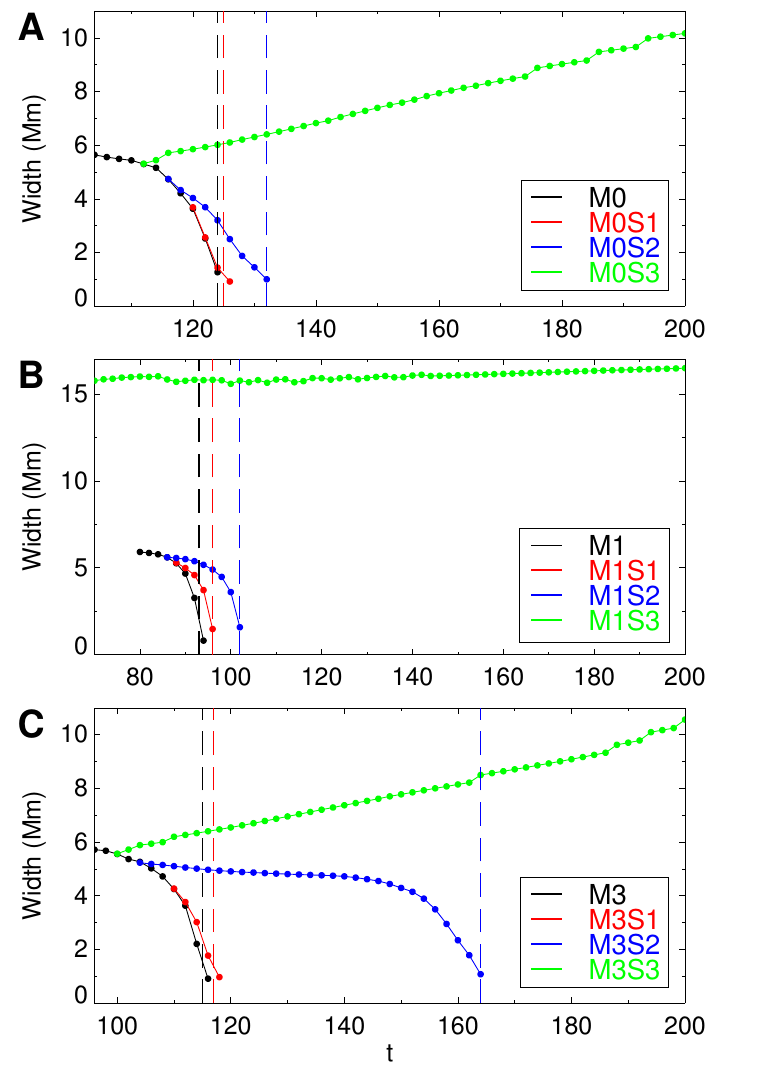}
	\caption{Temporal evolution of the thickness of current layer that formed in the core field in nine driving-stopped simulations. \textbf{A}, \textbf{B}, and \textbf{C} represent simulation M0, M1, and M3, respectively. The details are similar to those in \Fig~\ref{fig:Jc_core_all}C. The thickness of the current layer is defined as the thickness at the maximum value of $J/B$ or the thickness at the thinnest part of the current layer morphology. 
	Note that in panel \textbf{B}, the core current layer is very thick before $t=80$, so the number of grid nodes used for Gaussian function fitting in M1 is different from that in M1S3.}
	\label{fig:CS_S_all}
\end{figure*}

\begin{figure*}[!htbp]
	\centering
	\includegraphics[width=0.8\textwidth]{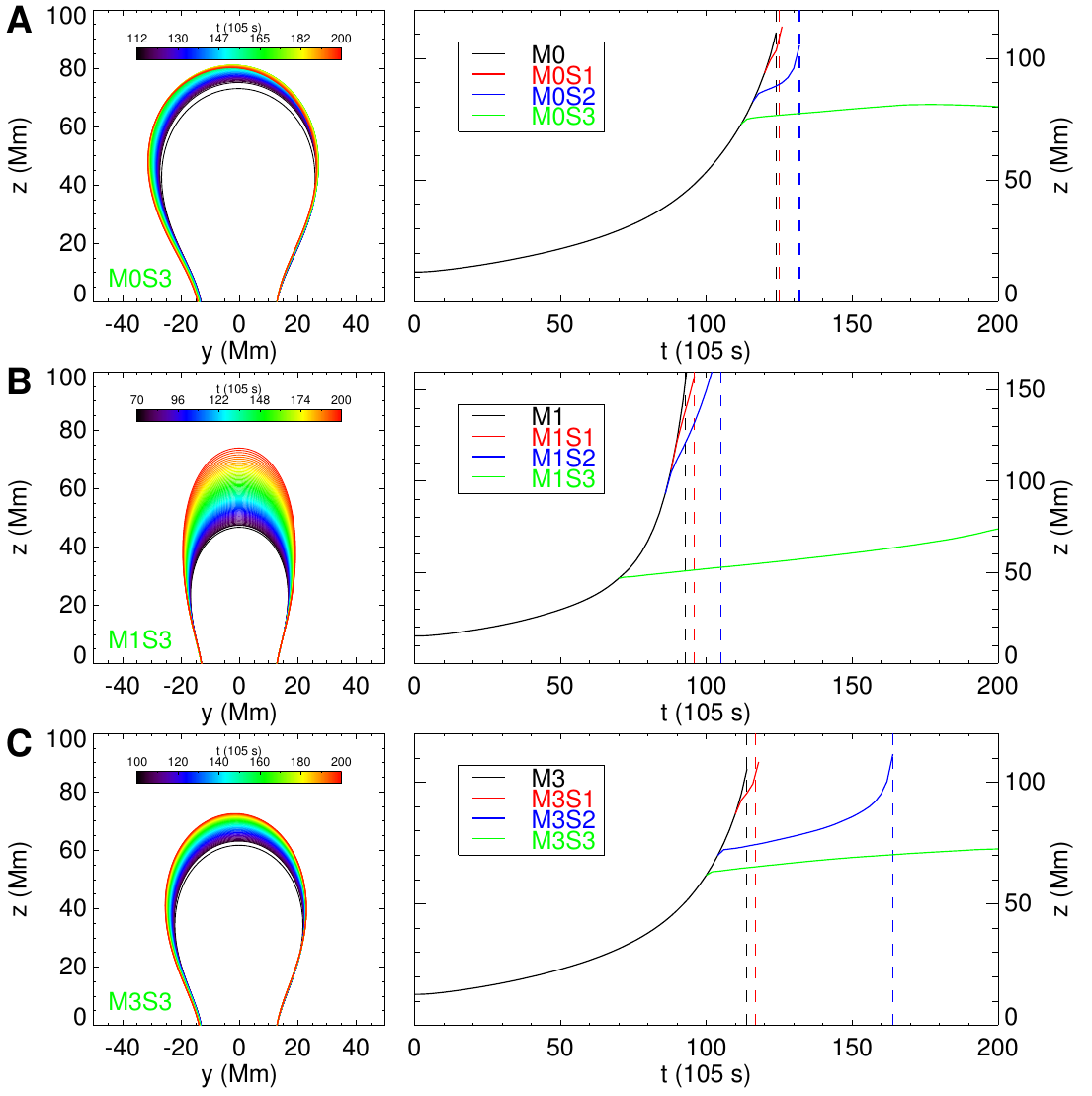}
	\caption{Evolution of the specific magnetic field line and its height with time in nine driving-stopped simulations. \textbf{A}, \textbf{B}, and \textbf{C} represent M0, M1, and M3, respectively. The details are similar to those in \Fig~\ref{fig:fieldlines}.}
	\label{fig:fieldlines_S}
\end{figure*}

\Fig~\ref{fig:energy_all_S} show the temporal evolution of energies in these driving-stopped simulations, where the red, blue, and green curves show respectively the results with the three incremental earlier moments of stopping the bottom driving (as denoted by S1, S2, and S3, respectively).
In all the three cases with different magnetograms, the S1 simulations produce eruption within a short time after stopping the driving, and the eruption starts only slightly later than that in the continuously-driven simulation. 
Similarly, all S2 simulations also produce eruption that starts somewhat later than those of original simulations. We also show the temporal evolution of thickness of core current layer in \Fig~\ref{fig:CS_S_all}. As can be seen, all the eruptions in S1 and S2 simulations start once the thickness of the core current layer decreases to close to the grid resolution. 
\Fig~\ref{fig:fieldlines_S} also shows the temporal evolution of the magnetic field line with same footpoint in the nine driving-stopped simulations, as shown in \Fig~\ref{fig:fieldlines}. 
The rising speed of this magnetic field line in all the simulations is slower than that of the original simulation at the moment of stopping driving.

\begin{figure*}[htbp]
	\centering
	\includegraphics[width=0.8\textwidth]{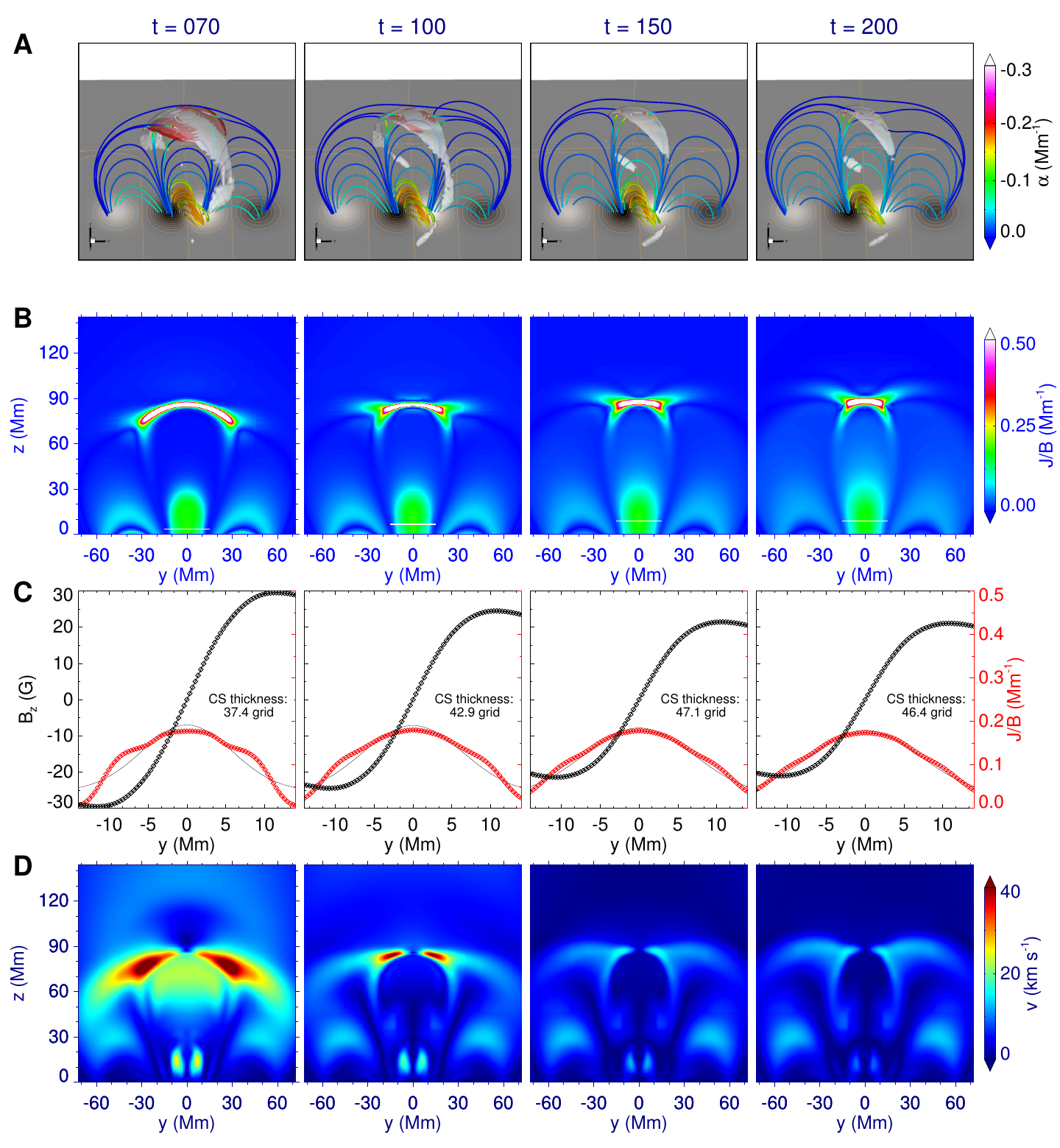}
	\caption{Evolution of breakout CS and velocity in driving-stopped simulation M1S3. \textbf{A}, 3D perspective view of magnetic field lines and structure of the breakout CS. The white iso-surface represent the current sheet with $J/B=0.5$, and the red iso-surface represent the Alfv$\acute{\text{e}}$nic Mach number with $v/v_{A}=1.2$.
	\textbf{B}, Normalized current in the $x=0$ slice.
	\textbf{C}, One dimensional profile of thickness of core current layer. 
	\textbf{D}, Velocity in the $x=0$ slice. (An animation of this figure is available. The animation is only of panels A-C.)} 
	\label{fig:fast_reconnection}
\end{figure*}

In contrast, all the S3 simulations do not produce eruption, and here we study why they fail to erupt despite the fact that the fast breakout reconnection has already began. 
\Fig~\ref{fig:fast_reconnection} (and its attached animation) shows evolutions of the breakout CS, core current distribution and velocity in the M1S3 simulation (and M0S3, M3S3). 
After stopping the driving, the breakout CS gradually shrank, and meanwhile the fast reconnection outflow also decayed.
Eventually the system relaxed to an equilibrium with residual velocity of around $10$~km~s$^{-1}$ at $t=200$, and the total magnetic energy is slightly dissipated by only $2.8$ percent in the long relaxation process without impulsive decrease (i.e., eruption).
In the final equilibrium, the breakout CS is still present due to upward pressing effect of the already sheared arcade below, but the reconnection outflow velocity has dropped to very low levels, which indicates that this breakout reconnection cannot self maintain if without the driving.
With relaxation of the system, the thickness of the core current layer slowly increases, i.e., a reversed process of CS formation.
During this process, although the core field also shows expansion, its speed is much slower that in the simulations with successful eruption (\Fig~\ref{fig:fieldlines_S} and animation of \Fig~\ref{fig:fast_reconnection}).   
	
Through the S3 simulations, we find that the eruption does not occur because the core CS in the system cannot form, although the fast breakout reconnection has already started. Therefore, our simulation suggests that the fast breakout reconnection cannot be used as a sign of the eruption onset. Moreover, the simulations M0S1 and M0S2 show that in the absence of breakout reconnection, the core CS can still form after stopping the driving, which indicates that breakout reconnection is not the necessary condition for triggering eruption.

When the system is driven by the bottom surface flow, the core field expand upward continuously and its current distribution is squeezed gradually (i.e., the thickness of the core current layer decreases).  
However, when the driving stops, both the squeezing speed and the expansion speed gradually weaken with relaxation of the magnetic field as aided by the small momentum viscosity. 
When these two speeds are dissipated and die before the core CS is formed, the eruption does not occur, while on the other hand, if they can survive until the core CS is formed, then the eruption can be triggered. 
That is, the formation of core CS is the key to the eruption.
The breakout reconnection does partly drive the formation of core CS, but the effect appears to be less important than that of the bottom driving in our simulations.

\subsection{The slow rise phase}
\label{subsec:slow_rise}
Although our simulations show that the formation of core CS is the key factor for initiation of eruption, it does not reveal at what stage or after what features the eruption is inevitable even without the bottom driving. All the S1 and S2 simulations form core CS and produce an eruption. 
Is it because the residual velocity, i.e., the inertia effect, dominates the evolution in those simulations and thus successfully buildup the core CS? Or because the system has passed a critical point after which the system is no more quasi-static in nature and the eruption is inevitable? 
To avoid the influence of residual velocity, we use an artificial frictional force to constrain the velocity of the system in the driving-stopped simulations. Specifically, we add an artificial friction term to the momentum equation, which is given as:
\begin{equation}\label{eq:momentum}
	\vec F = -f \frac{\rho \vec v}{t_A}
\end{equation}
where $f$ is an adjustable coefficient and $t_A$ is the local Alfv$\acute{\text{e}}$nic time, namely, $t_A = \frac{1}{v_A} = \frac{\sqrt{\rho}}{B}$. This frictional term means that the velocity is dissipated to zero within the time of $\frac{1}{f}t_{A}$. We note that such friction does not exist in the real corona, while here it is used only for examining the effect of the inertia of the system in the initiation of the eruption. If the MHD system evolves in a quasi-static nature, the frictional force can help it relax to an equilibrium quickly, and therefore has been frequently used in coronal force-free field reconstructions~\citep{Valori2007, Jiang2013b, Guo2016}.
	
For comparison, we select two simulations, M0S2 and M1S2, and use three different values of $f=0.1$, $0.3$, and $1$, respectively, which represents a small, median, and large frictional force. Furthermore, the continuously-driven simulations M0 and M1 are also run with frictional force of $f=0.3$, for the aim to show how the frictional force affects the eruption behavior as compared to those with no frictional force. See Table~\ref{tab:simulation_M0M1} for specific friction setting and eruption onset.

\begin{table}[!htbp]
	\centering
	\begin{tabular}{cccc}
		\hline\hline
		\multirow{2}{*}{Map} & Eruption  & Simulation & Eruption Onset \\
		& Onset & (with friction) & (with friction)  \\
		\hline
		\multirow{4}{*}{M0} & \multirow{4}{*}{123} &  M0 ($f=0.3$) & 128 \\
		&&M0S2 ($f=0.1$)&135 \\
		&&M0S2 ($f=0.3$)&140 \\
		&&M0S2 ($f=1$)  &160 \\
		\hline
		\multirow{4}{*}{M1} & \multirow{4}{*}{93}  &  M1 ($f=0.3$) & 96.5 \\
		&&M1S2 ($f=0.1$)&105.5 \\
		&&M1S2 ($f=0.3$)&111 \\
		&&M1S2 ($f=1$)  &127 \\
		\hline
	\end{tabular}
	\caption{The definition of eight simulations with friction.}
	\label{tab:simulation_M0M1}
\end{table}

\begin{figure*}[!htbp]
	\centering
	\includegraphics[width=0.6\textwidth]{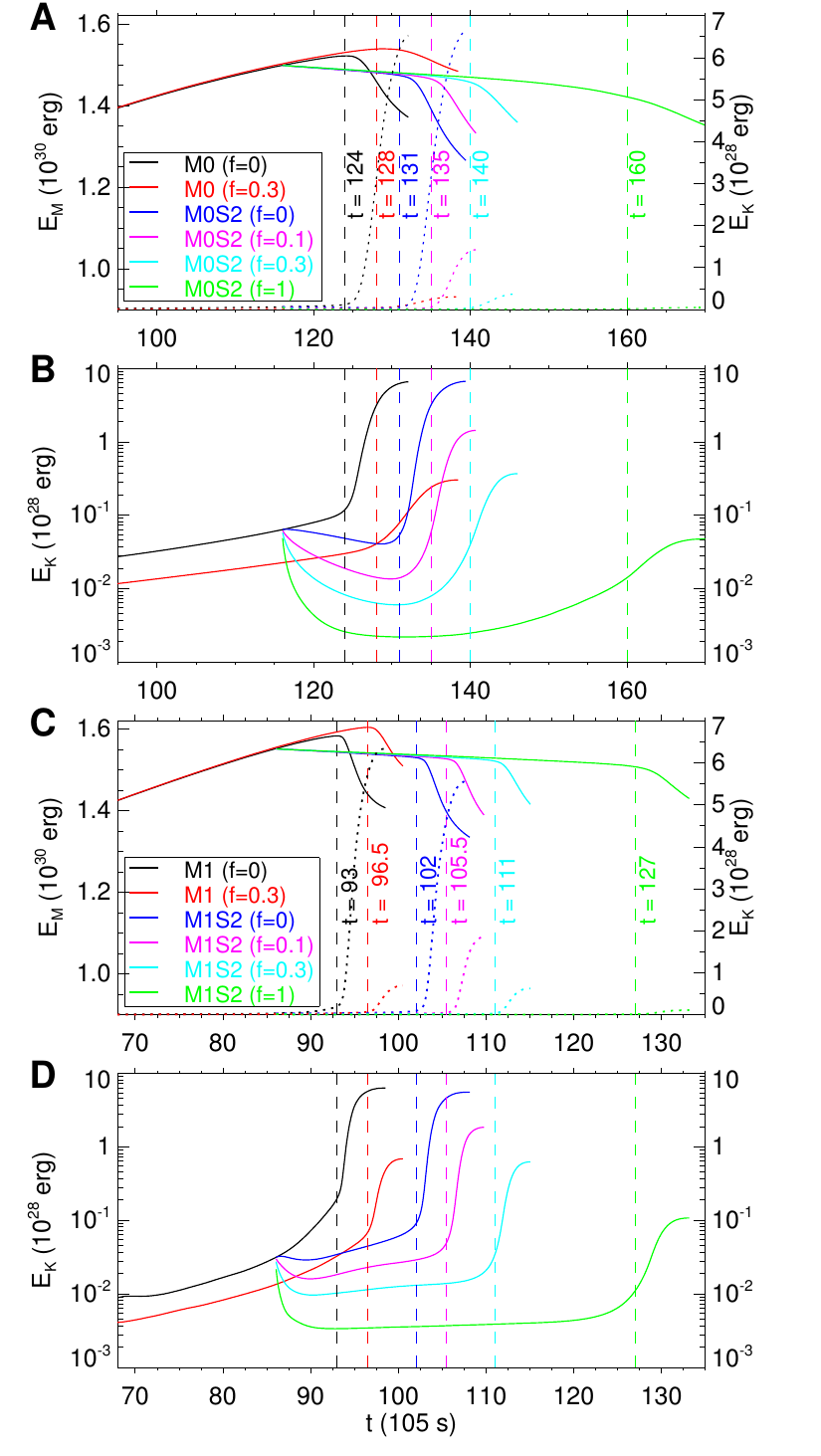}
	\caption{Temporal evolution of energies in simulations with friction. \textbf{A}, Evolution of magnetic energy $E_{\rm M}$ (solid lines) and kinetic energy $E_{\rm K}$ (dotted lines) in M0 simulations with different friction. \textbf{B},  Evolution of kinetic energy but on the logarithmic axis. The vertical dashed lines with the time numbers show the onsets of the eruption. \textbf{C} and \textbf{D} have the same details as \textbf{A} and \textbf{B}, respectively, but shown for M1 simulations. (An animation of this figure is available. The animation contains the same diagrams as panel C in \Fig~\ref{fig:Breakout_slice} and the evolution of kinetic energies in this \Fig.)}
	\label{fig:energy_M0M1}
\end{figure*}

\begin{figure*}[!htbp]
	\centering
	\includegraphics[width=0.8\textwidth]{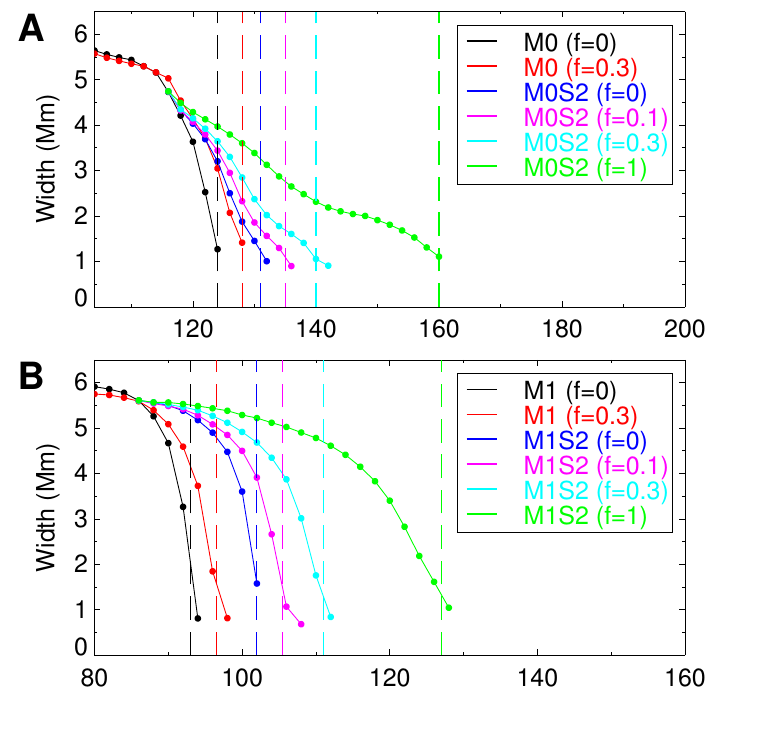}
	\caption{Temporal evolution of the thickness of current layer that formed in the core field in simulations with friction. \textbf{A} and \textbf{B} represent simulations M0 and M1, respectively. The details are similar to those in \Fig~\ref{fig:Jc_core_all}C.}
	\label{fig:CS_M0M1}
\end{figure*}

\begin{figure*}[!htbp]
	\centering
	\includegraphics[width=0.8\textwidth]{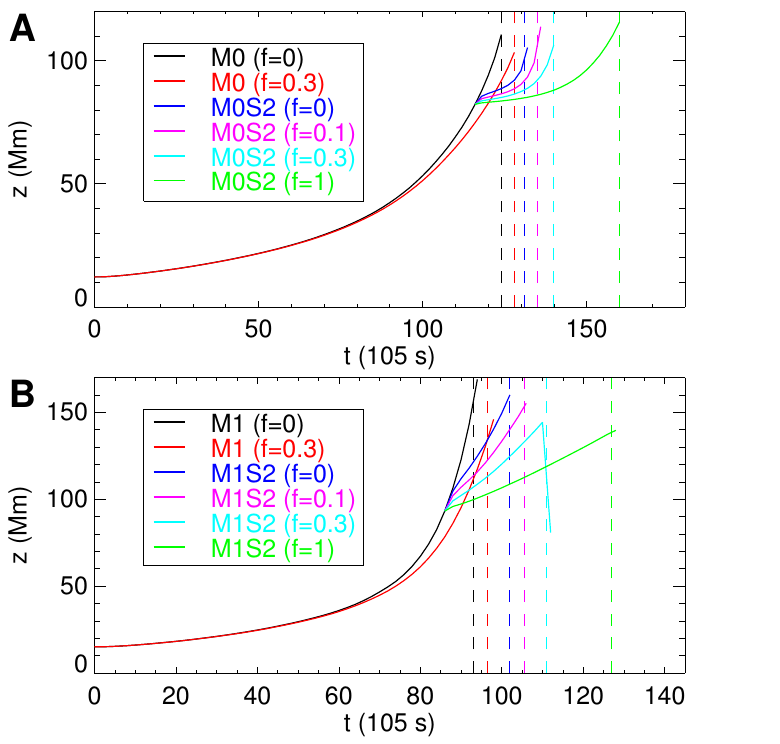}
	\caption{Evolution of the specific magnetic field line and its height with time in simulations with friction. \textbf{A} and \textbf{B} represent simulations M0 and M1, respectively. The details are similar to those in \Fig~\ref{fig:fieldlines}.}
	\label{fig:fieldlines_M0M1}
\end{figure*}

\Fig~\ref{fig:energy_M0M1} shows evolution of energies in the eight simulations with friction and compared with those four cases without friction. With the friction, the residual kinetic energy in all the M0S2 and M1S2 simulations decays quickly and substantially, as shown clearly by the logarithmic scale of the kinetic energy in \Fig~\ref{fig:energy_M0M1}B and D. Therefore the inertia effect in driving the evolution is largely removed. However, it can be seen that, although different values of friction give different results, all simulations produce eruption, i.e., a rapid release of magnetic energy, and formation and rise of the MFR as driven by the reconnection (see animation of \Fig~\ref{fig:energy_M0M1}).
Due to the applied friction, the onset of eruption is delayed, and the greater the friction coefficient, the later the onset of eruption.
It is obvious that the application of friction reduces the velocity of the evolution in the system, including the thinning speed of core current layer (\Fig~\ref{fig:CS_M0M1}) and the expansion speed of the core field (\Fig~\ref{fig:fieldlines_M0M1}).

For instance, in the simulation with the very large friction coefficient $f=1$, the eruption also occurs, despite that the rates of magnetic energy release and kinetic energy increase are much weaker than those with smaller friction. This is because the strong friction slows down significantly the eruptive process of the system, but it cannot prevent the occurrence of eruption.
In the evolution of the core current layer as shown in \Fig~\ref{fig:CS_M0M1} (green curves), it can still be seen that the thickness of current layer continuously decreases to the grid resolution, though at a low thinning speed, and finally triggers magnetic reconnection. During this process, the core field also expands continuously at a low speed (\Fig~\ref{fig:fieldlines_M0M1}).
	
Combining these simulation results of M0S2 and M1S2 with friction, we find that the eruption becomes inevitable after the system evolves to a certain moment, regardless of whether it is a single bipolar field or a quadrupolar field.
Certainly, this moment is close to the onset of eruption.
We consider that the system at this moment has entered a slow rise phase which is no more quasi-static since the friction cannot settle down it. The slow-rise phase is often observed in filament eruptions for a short period (of a few to tens of minutes) in which the filament shows a slow rise with a speed faster than that of the quasi-static phase but significantly slower than that of the eruption phase~\citep[e.g.,][]{Cheng2020}. 
Once the system is driven to enter this stage, it will always evolve spontaneously to produce an eruption, regardless of whether there is still a bottom driving.
This is independent of the residual velocity in the system.

\section{Conclusion} \label{sec:con}
The fundamental mechanism of solar eruption initiation, in which a CS slowly forms within the sheared core and reconnection in the CS triggers and drives an eruption, can be applied to the bipolar magnetic field that exists universally on the Sun~\citep{Jiang2021b, Bian2022a, Bian2022b}, but most violent solar eruptions often originate from complex sunspot group that consist of multipolar magnetic fields, in particular, the $\delta$ sunspot groups~\citep{Kuenzel1959, Guo2014, Yang2017a, Toriumi2019, Toriumi2021}. In this paper, we extended this fundamental mechanism to a typical multipolar magnetic field: a quadrupolar magnetic configuration containing a null point above the core field, which is the basic configuration of the classical breakout model. 

To this end, we have carried out a series of fully 3D MHD simulations with comparative analysis. These simulations all have the same core bipolar field but different background fields (and with a reference case without background field), so they have magnetic null point at different heights. 
By continuously shearing (i.e., energizing) the core field at the bottom surface, all these simulations show magnetic evolution to eruption in a similar process, which, in the point of view of magnetic topology, is identical to the classical breakout model. Initially, the core field expand upwards, squeezing the null point into a horizontal CS with breakout reconnection subsequently occurring. Meanwhile a vertical CS gradually forms within the sheared core field, and finally the eruption is triggered with a twisting flux rope expelled out from the core field. The evolutions of the magnetic and kinetic energies of the system indicate clearly the eruption onset only begins when reconnection starts in the core CS. Comparison of the different simulations show that the lower the height of null point is, the earlier the breakout reconnection starts, and also the earlier the eruption onset is. This is because that the breakout reconnection can help the formation of core CS by letting the core field expand faster. Nevertheless, the fast breakout reconnection cannot lead the core field into a dynamic evolving phase, and the thinning of the core current layer still proceeds in a slow way until the eruption. 

To pin down the key factor in initiation of the eruption, we carried out controlled experiments by stopping the bottom driving at particular moments that are after the beginning of fast breakout reconnection but prior to the formation of the core CS. These experiments show that the system cannot produce eruption if the core CS fails to form (as the bottom driving is stopped too early), even though the fast breakout reconnection has started. Furthermore, the fast breakout reconnection cannot be self-maintained if without the driving, therefore not able to establish a positive feedback-loop between the breakout reconnection and the expansion of the core field. Thus, our simulation suggests that the key to eruption initiation in such multipolar configuration remains to be the slow formation of the CS in the sheared core rather than the onset of fast breakout reconnection.

A further set of experiments with frictional forces shows that the eruption is inevitable after the system enters a slow rise phase. The physical nature of the system at this phase is different from the quasi-static evolution since, even with a strong friction applied, the system still expands and forms a core CS, although with a reduced speed. Such behavior exists in both the single bipolar field and the quadrupolar field, and thus is not dependent on the breakout reconnection. Observations of filament eruptions seem to show that the slow rise phase can be clearly distinguished by fitting the height-time curve, as well as the velocity-time and acceleration-time curve~\citep{Zhang2001, Zhang2004, Fan2019, Cheng2020, Liu2021}. A future study will be performed to study the physical nature of the system at the slow rise phase and to find suitable parameters to identify whether the system enters this stage.

In summary, through comparative analysis of a series of full 3D MHD simulations of eruption initiated within a quadrupolar magnetic configuration, we show that the breakout reconnection indeed helps the formation of the core CS by letting the core field to expand faster, but the eruption cannot occur when the bottom surface driving is stopped well before the core CS is formed, even though fast reconnection has already been triggered in the breakout CS. The breakout reconnection alone seems not able to establish a positive feedback loop between itself and the expansion of the core field (that leads to a fast formation of the core CS). 
This study clarified the role of breakout reconnection, and confirmed formation of the core CS as the key to the eruption initiation in a multipolar magnetic field, which is consist with the fundamental mechanism of solar eruption in our previous series of bipolar field simulations~\citep{Jiang2021b, Bian2022a, Bian2022b, Bian2022, Wang2022c}.

\begin{acknowledgments}
\modulolinenumbers[10]
This work is jointly supported by National Natural Science Foundation of China (NSFC 42174200), the Fundamental Research Funds for the Central Universities (HIT.OCEF.2021033), Shenzhen Science and Technology Program (RCJC20210609104422048), and Shenzhen Technology Project JCYJ20190806142609035. The computational work was carried out on TianHe-1(A), National Supercomputer Center in Tianjin, China and the ISSAT Cluster computing system (HIT, Shenzhen).
\end{acknowledgments}

\bibliography{mylab}
\bibliographystyle{aasjournal}

\end{document}